\documentclass[pra,aps,twocolumn,10pt]{revtex4-1}
\usepackage{graphicx}

\begin{document}

\title{Maximum Entropy, Word-Frequency, Chinese Characters, and Multiple Meanings}

\author{Xiaoyong Yan$^{1,2}$}
\author{Petter Minnhagen$^{3}$}
\email{ Petter.Minnhagen@physics.umu.se}

\affiliation{
$^{1}$Systems Science Institute, Beijing Jiaotong University, Beijing 100044, China\\
$^{2}$Big Data Research Center, University of Electronic Science and Technology of China, Chengdu 611731, China\\
$^{3}$IceLab, Department of Physics, Ume{\aa} University, 901 87 Ume{\aa}, Sweden
}

\begin{abstract}
The \textit{word}-frequency distribution of a text written by an author is well accounted for by a maximum entropy distribution, the RGF (random group formation)-prediction. The RGF-distribution is completely determined by the \textit{a priori} values of the total number of words in the text ($M$), the number of distinct words ($N$) and the number of repetitions of the most common word ($k_{max}$). It is here shown that this maximum entropy prediction also describes a text written in \textit{Chinese characters}. In particular it is shown that although the same Chinese text written in \textit{words} and \textit{Chinese characters} have quite differently shaped distributions, they are nevertheless \textit{both} well predicted by their respective three \textit{a priori} characteristic values. It is pointed out that this is analogous to the change in the shape of the distribution when translating a given text to another language. Another consequence of the RGF-prediction is that taking a part of a long text will change the input parameters ($M, N, k_{max}$) and consequently also the shape of the frequency distribution. This is explicitly confirmed for texts written in Chinese characters.
Since the RGF-prediction has no system-specific information beyond the three \textit{a priori} values ($M, N, k_{max}$), any specific language characteristic has to be sought in systematic deviations from the RGF-prediction and the measured frequencies. One such systematic deviation is identified and, through a statistical information theoretical argument and an extended RGF-model, it is proposed that this deviation is caused by multiple meanings of Chinese characters. The effect is stronger for Chinese characters than for Chinese words. The relation between Zipf's law, the Simon-model for texts and the present results are discussed.
\end{abstract}

\maketitle

\section*{Introduction}

The scientific interest in the information-content hidden in the frequency statistics of words and letters in a text goes at least back to Islamic scholars in the ninth century. The first practical application of these early endeavors seems to have been the use of frequency statistics of letters to decipher cryptic messages \cite{singh2000}. The more specific question of what \textit{linguistic} information is hidden in the \textit{shape} of the word-frequency distribution stems from the first part of the twentieth century when it was discovered that the words in a text typically have a broad ``fat-tailed" shape, which often can be well approximated with a power law over a large range \cite{estroup16,zipf32,zipf35,zipf49}. This led to the empirical concept of Zipf's law which states that the probability that a word occurs $k$-times in a text, $P(k)$, is proportional to $1/k^2$ \cite{zipf32,zipf35,zipf49}. The question is then what principle or property of a language causes this power law distribution of word-frequencies and this is still an ongoing research \cite{mand53,li92,baayen01,cancho03,mont01}. In the middle of the twentieth century Simon in \cite{simon55} instead suggested that since quite a few completely different systems also seemed to follow Zipf's law in their corresponding frequency distributions, the explanation of the law must be more general and stochastic in nature and hence independent of any specific information of the language itself. Instead he proposed a random stochastic growth model for a book written one word at a time from beginning to end. This became a very influential model and has served as a starting point for much later works \cite{kanter95,doro01,zanette05,masucii06,cattuto06,lu13}. However, it was recently pointed out that the Simon-model has a fundamental flaw: the rare words in the text are more often to be found in the later part of the text, whereas a real text is to very good approximation translational invariant: the first half of a \textit{real} text has, provided it is written by the same author, the same word-frequency distribution as the second \cite{bern10,bern11}. So, although the Simon-model is very general and contains a stochastic element, it is still history dependent and, in this sense, it leads to a less random frequency distribution than a real text. An extreme random model was proposed in the middle of the twentieth century by Miller in \cite{miller57}: the resulting text can be described as being produced by a monkey randomly typing away on a typewriter. The monkey book is definitely translational invariant, but its properties are quite unrealistic and different from a real text \cite{bern11b}. 

\begin{figure*}[!htbp]
\includegraphics[width=1\textwidth]{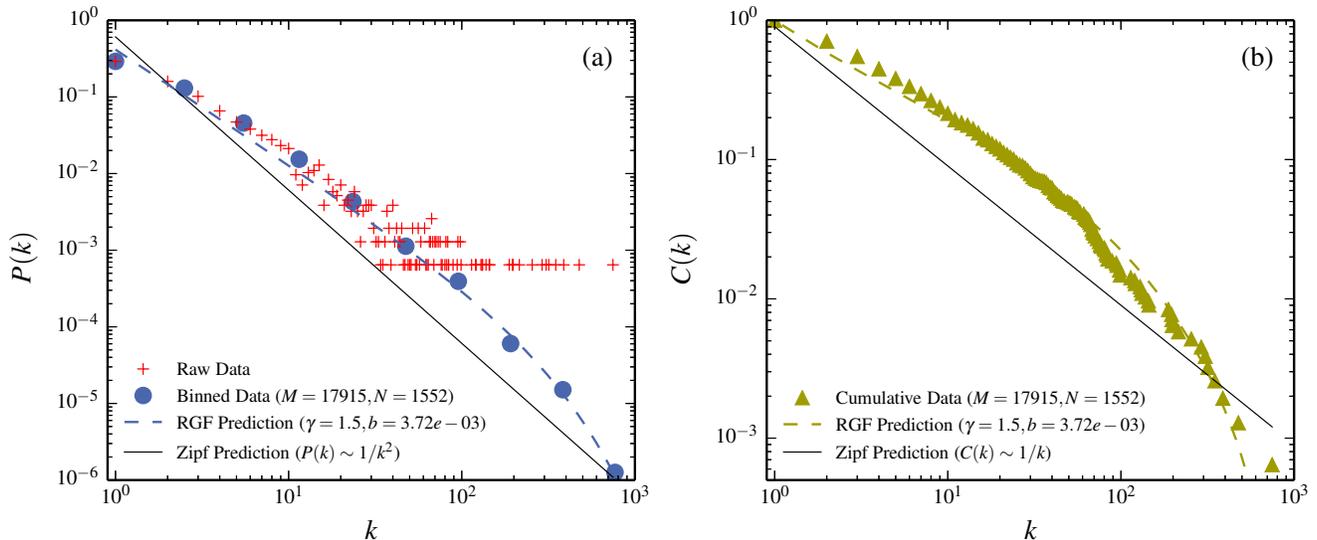}
\caption{
{\bf Frequency of Chinese characters for the novel \textit{A Q Zheng Zhuan} by Xun Lu and comparison with the RGF-prediction and the Zipf's law expectation.} (a) Compares the probability, $P(k)$, for a character to appear $k$-times in the text: crosses are raw data, filled dots are the log2-binned data, the straight line is the Zipf's law expectation, and the dashed curve is the RGF-prediction. The RGF predicts the dashed curve directly from the three values $(M,N,k_{max})$ (see Table \ref{tb:1} for the input values and the corresponding predicted output values from RGF).  (b) The same features in terms of the cumulative distribution $C(k)=\sum_{k'\geq k}P(k)$: filled triangles are the data, the straight line the Zipf's law expectation and the dashed curve the RGF-prediction. RGF gives a very good \textit{ab initio} description of the data which differs substantially from the Zipf's law expectation {(Note that the RGF-prediction is based solely on the raw data and predicts both the binned data in (a) and the cumulant data in (b))}.
}
\label{fig:AQ}
\end{figure*}

The RGF (random group formation)-model, which is the basis for the present analysis, can be seen as a next step along Simon's suggestion of system-independence \cite{baek11}. Instead of introducing randomness from a stochastic growth model, RGF introduces randomness directly from the maximum entropy principle \cite{baek11}. An important point of the RGF-theory is that it is predictive: if {the} only knowledge of the text is $M$ (total number of words), $N$ (number of distinct words), and $k_{max}$ (number of repetitions of the most common word), then RGF provides a complete prediction of the probability distribution $P(k)$. This prediction includes the functional form, which embraces Gaussian-like, exponential-like and power-law-like shapes; the form is determined by the sole knowledge of $(M, N, k_{max})$. A crucial point is that, if the maximum entropy principle, through RGF, gives a very good description of the data, then this implies that the values $(M,N, k_{max})$ incorporate all information contained in the distribution
 $P(k)$, which makes the prediction neutral and void of more specific characteristic features. More specific text information is, from this view-point, associated with systematic deviations from the RGF-prediction. 

Texts sometimes deviate significantly from the empirical Zipf's law and a substantial part of work has been devoted to explain such deviations. These explanations usually involve text- and language specific features. However, from the RGF point of view, such explanations appear rather redundant and arbitrary, whenever the RGF-prediction agrees with the data. This point of view has been further elucidated in \cite{bokma13} for the case of species divided into taxa in biology.

In a recent paper by L. L\"u \textit{et al.} \cite{lu13} it was pointed out that the character frequency-distribution for a text written in Chinese characters differs significantly from Zipf's law, as had also been noticed earlier \cite{zhao90,rous92,shtrik94,ha00}. {Chinese characters carry specific meanings. For example, `hu\'{i}' and `ji\={a}' are two Chinese characters carrying the elementary meanings of ``return" and ``home", respectively. In general a Chinese character can also carry multiple meanings, where the relevant meaning has to be deduced from the context. A Chinese word corresponds to one, two or more characters, \textit{e.g.} the two characters `hu\'{i}' and `ji\={a}' can be combined into the Chinese word `hu\'{i},ji\={a}' denoting the concept of ``returning home". Thus both Chinese characters and Chinese words carry meanings which can be single or multiple. Roughly a word in Chinese corresponds to about 1.5 characters on the average and typically more than 90\% of the words in a novel are written with one or two characters, where about 50\% of the words are written by one character and 40\% with two. The remaining ones are made up of more than two Chinese characters.}

{The Chinese character frequency distribution } is illustrated in Fig. \ref{fig:AQ}. The straight line in the figure is the Zipf's law expectation. From a Zipf's law perspective one might then be tempted to conclude that the deviations between the data and Zipf's law have something to do specifically with the Chinese language or the representation in terms of Chinese characters, or perhaps a bit of both. However, the dashed curve in the figure is the RGF-prediction. This prediction is very close to the data, which suggests that beyond the three characteristic numbers $(M,N,k_{max})$ [total number of Chinese characters, distinct characters, and the number of repetitions of the most common character] there is \textit{no} specifically Chinese feature, which can be extracted from the data.

A crucial point for reaching our conclusions in the present paper is the distinction between a predictive model like RGF and conventional curve-fitting. This can be illustrated by Fig. \ref{fig:AQ}(b): if your aim is to fit the lowest $k$-data points in Fig. \ref{fig:AQ}(b) ({\it e.g.} $k=1$ to 10) with an {\it ad hoc} two parameter curve you can obviously do slightly better than the dashed curve in the Fig. \ref{fig:AQ}(b). However, the dashed curve is a {\it prediction} solely based on the knowledge of the right-most point in Fig. \ref{fig:AQ}(b) ($k_{max}=747$) and the average number of times a character is used ($M/N=11.5$). RGF predicts where the data points in the interval $k=1-10$ in Fig. \ref{fig:AQ}(b) should fall  {\it without} any explicit a priori knowledge of their whereabouts and with very little knowledge of anything else. This is the crucial difference between a prediction from a model and a fitting procedure and this difference carries over into the different conclusions which can be drawn from the two procedures. Another illustration is the fact that although the data in  Fig. \ref{fig:AQ}(b) cannot be described by a Zipf's-line with slope -1, such a line can be fitted to the data over a narrow range somewhere in the middle. Such an {\it ad hoc} fitting has no predictive value.

Specific information about the system may be reflected in deviations from the RGF-prediction \cite{bokma13}. One such possible deviation is discussed. It is also suggested that the cause of this deviation is multiple meanings of Chinese characters. A statistical information based argument for this conclusion is presented together with an extended RGF-model.

The method section starts with a brief recapitulation of the RGF-theory, as well as the Random Book Transformation, which allows for the analysis of sub-parts of the novels. Both these methods are used as starting points when analyzing the frequency distribution for two Chinese novels. The Chinese character-frequency distributions are compared to the corresponding word-frequency distributions for both novels, as well as for parts of the novels. The results from these comparisons lead to an information theory which makes it possible to approximately include the multiple meanings of Chinese characters. It is pointed out that the existence of words with multiple meanings isn't a characteristic specific to Chinese, but a general feature of languages. The frequency distribution of the elementary entities of a written language (words or characters) is therefore influenced by the distribution of meanings over these entities, in a characteristic way. Conclusions are discussed in a last section.

\section*{Methods}

\subsection*{Random Group Formation}

The random group formation describes {the }general situation {in which }$M$ objects are randomly grouped together into $N$ groups \cite{baek11}. The simplest case is when the objects are denumerable. Then if you know $M$ and $N$ the most likely distribution of group sizes, $N(k)$ (number of sizes with $k$ objects), can be obtained by minimizing the information average $I[N(k)]=N^{-1}\sum N(k)\ln(kN(k))$ with respect to the functional form of $N(k)$, subject to the two constraints that $N^{-1}\sum N(k)k=<k>=M/N$ and $\sum N(k)=N$. Note that the information to localize an object in one of the groups of size $k$ is $\log_2(kN(k))$ in bits and $\ln(kN(k))$ in nats. Minimizing the average information $I[N(k)]$ is equivalent to maximizing the entropy \cite{baek11}. Thus RGF is a way to apply the maximum entropy principle to this particular class of problems. The result of the simplest case is the prediction $N(k)=A\exp(-bk)/k$ \cite{baek11}. However, in more general cases there might be many additional constraints and in addition all the objects might not lend themselves to a simple denumerization. The point is that in many applications you \textit{do} know that there must be additional constraints relative to the simplest case \textit{but} you have no idea what they might be. The RGF-idea is then based on the observation that any deviation from the simplest case will be reflected in a change of the entropy $S[N(k)]=-\sum_k N(k)/N \ln(kN(k)/N)$. This can then be taken into account by incorporating the actual value of the entropy $S$ as an additional constraint in the minimizing of $I[N(k)]$. The resulting more general prediction then becomes $N(k)=A\exp(-bk)/k^\gamma$ \cite{baek11}. Thus RGF transforms the three values $(M,N,S)$ into a complete prediction of the group-size distribution. This also means that the form of the distribution is determined by the values $(M,N,S)$ and includes a Gaussian limit (when $\gamma=(M/N)b$ and $(M/N)^2/\gamma$ is small), exponential (when $\gamma=0$), power-law (when $b=0$) and anything in between.

In comparison with earlier work, one may note that the functional form $P(k)=A\exp(-bk)/k^\gamma$ has been used before when parameterizating distributions as described {\it e.g.} by Clauset {\it et al} \cite{clauset09}  and that such a functional form can obtained from a maximum entropy as described {\it e.g.} by Visser \cite{visser13}. The difference with our approach is the   connection to minimal information which opens up the predictive part of the RGF. {As emphasized in the Introduction, it} is this predictive aspect which is crucial in  {the present} approach and which lends itself to the generalization of including multiple meanings of characters.

The RGF-distribution was in \cite{baek11,bokma13,lee12,baek11b} shown to apply to a variety of systems like words in texts, population in counties, family names, distribution of richness, distribution of species into taxa, node sizes in metabolic networks, etc. In case of words, $N$ is the number of different words, $M$ is the total number of words, and $N(k)$ is the number of different words which appears $k$ times in the text. In English the largest group consists of the word ``the" and its occurrence in a text written by an author is a statistically very well defined: it is typically about 4\% of the total number of words \cite{bern11,baek11}. As a consequence one may replace the three values $(M,N,S)$ by the three values $(M,N,k_{max})$. Both choices completely determine the parameters $(A,b,\gamma )$ in the RGF-prediction. However, the latter choice has the {practical }advantage that $k_{max}$, the number of repetitions of the most common word, is more directly accessible and statistically very well-defined \cite{footnote}. For example, if $k_{max}$ is close to the average $<k>=M/N$, such that $(k_{max}-<k>)/<k><<1$ then the RGF-prediction approaches a Gaussian, which comes as no surprise because a Gaussian is just the outcome of the maximum entropy principle for such a narrow distribution \cite{baek11}.

\subsection*{Random Book Transformation}

In general, the distribution for a system, which falls into the RGF-class, has a distribution with a shape which depends on $M$. Since $M$ for a text is the total number of words, this means that the frequency distribution is text-length dependent. The reason is that if you start from a text characterized by $(M,N,k_{max})$,  then the corresponding value for a half of the text is characterized by $(M_{1/2},N_{1/2}, k_{max_{1/2}})$. Here $M_{1/2}=M/2$ by definition, $k_{max_{1/2}}=[k_{max}]/2$ because the most common word is to good approximation equally distributed within the text, but $N_{1/2}$ is non trivial.  {In the present investigation we need a method to separate between changes in the frequency distribution due to multiple meanings and due to the size of the text. For this purpose we use the Random Book Transformation (RBT) discussed in \cite{bern10}, where it was shown that the text-length dependence of the average $N$, when taking a part of a given text, is to good approximation a neutral feature: it is to good approximation the same as when you randomly delete the corresponding amount of words from the text.} The process of changing the length of a text by randomly deleting words is a simple statistical process which transforms the probability distribution $P_M(k)=N(k)/N$ for the full text into $P_{M/n}(k)$ for the n$^{th}$ part of the text by 
\begin{equation}
\label{eq_RBT1}
\textbf{P}_{M/n}(k)=B\sum_{k'=k}^M \textbf{A}_{kk'}\textbf{P}_M(k'),
\end{equation}
where $\textbf{P}_{M/n}$ and $\textbf{P}_{M}$ are column matrices corresponding to $P_{M/n}$ and $P_{M}$. The transformation matrix $\textbf{A}_{k'k}$ is given by 
\begin{equation}
\label{eq_RBT2}
\textbf{A}_{kk'}=(n-1)^{k'-k}n^{k'}C^{k'}_k,
\end{equation}
where $C^{k'}_k$ is binomial coefficient. $B$ is given by the normalization condition
\begin{equation}
\label{eq_RBT3}
B^{-1}=\sum_k^M\sum_{k'=k}^M \textbf{A}_{k'k}\textbf{P}_M(k').
\end{equation}

As shown in the next section, this simple random book transformation also to good approximation applies to text written in Chinese characters.

\begin{figure*}[!htbp]
\includegraphics[width=1\textwidth]{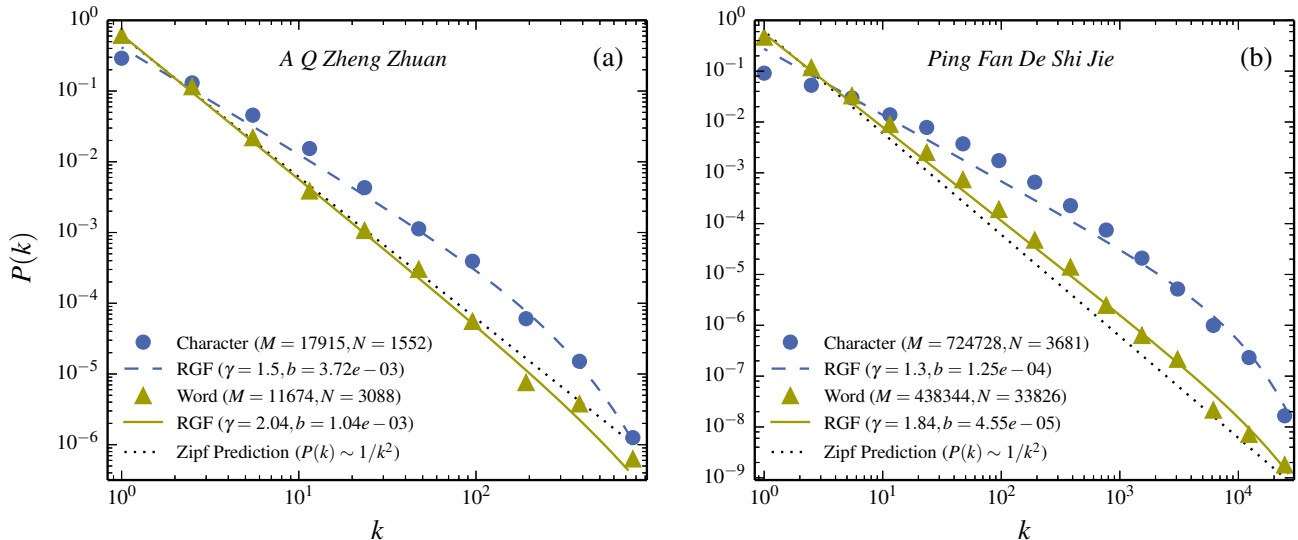}
\caption{
{\bf Comparison between Chinese texts in characters and words.} (a) Comparison between characters and words for the novel \textit{A Q Zheng Zhuan} by Xun Lu together with the respective RGF-predictions. (b) The same comparison for the novel \textit{Ping Fan De Shi Jie} by Yao Lu. Filled dots correspond to the binned data for Chinese characters and filled triangles the data for words. Full and dashed curves correspond to the respective RGF-predictions and dotted straight lines are the Zipf's law expectations for the word-frequency distribution. The respective ``state"-variables $(M,N,k_{max})$ and the corresponding RGF-predictions are given in Table  \ref{tb:1}. Note that the translation between words and characters is a deterministic process. Yet the ``state"-variables $(M,N,k_{max})$ suffice to predict the change in frequency distribution caused by the translation between words and characters.
}
\label{fig:word_character}
\end{figure*}

\begin{figure*}[!htbp]
\includegraphics[width=1\textwidth]{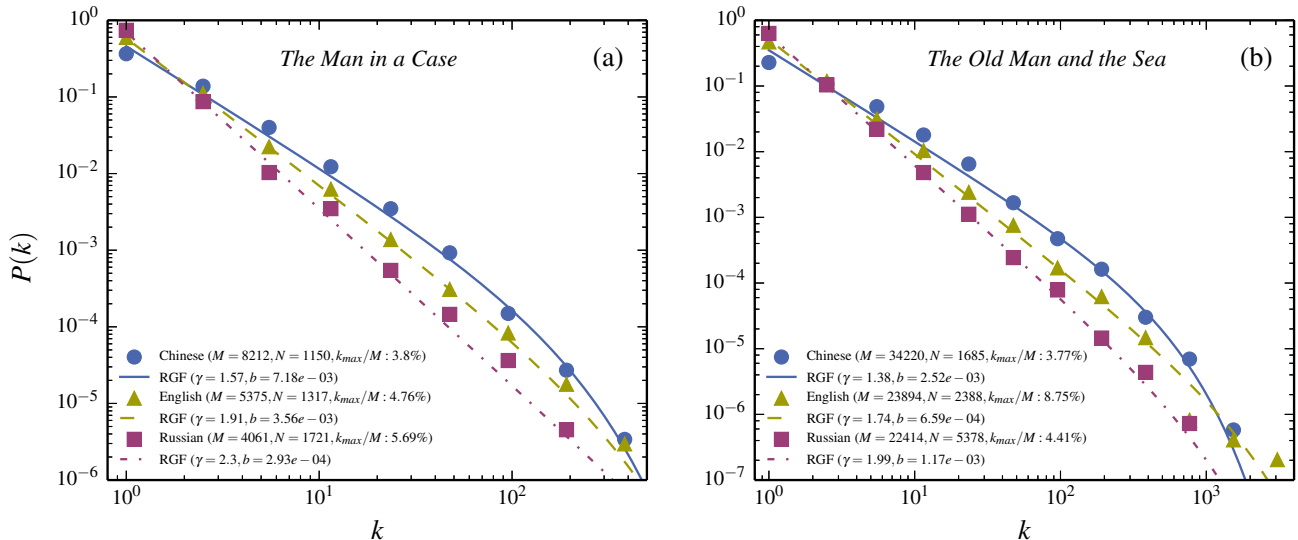}
\caption{
{ {\bf Similarity of translation between words and characters versus words of different languages.} (a) The Russian short story \textit{The Man in a Case} by A. Chekhov, and its translations into English words and Chinese characters (triangles, squares, and filled dots, respectively). The RGF-predictions are given by the curves (dashed dotted, dashed, and full, respectively, The RGF-prediction completely characterizes a frequency distribution in terms of the total number of words/characters ($M$), the number of specific words/characters ($N$), and how many of the total number of words/characters  are given by the most common word/character ($k_{max}/M$). Each such triple ($M$, $N$, $k_{max}$) gives a unique prediction-curve [($M$, $N$, $k_{max}$)=(4061, 1721, 231), (5375, 1317, 256), and (8212, 1150, 312), respectively]. The agreement shows that words and characters are entirely analogous with respect to frequency distributions. (b) illustrates the same thing starting from the English novel \textit{The Old Man and the Sea} and translating into Russian words and Chinese characters. The triples are this time ($M$, $N$, $k_{max}$)=(22414, 5378, 988), (23894, 2388, 2091), and (34220, 1685, 1289), in the order Russian, English and Chinese characters (Data points and RGF-curves, as in (a)).
}}
\label{fig:trans}
\end{figure*}

\begin{figure*}[!htbp]
\includegraphics[width=1\textwidth]{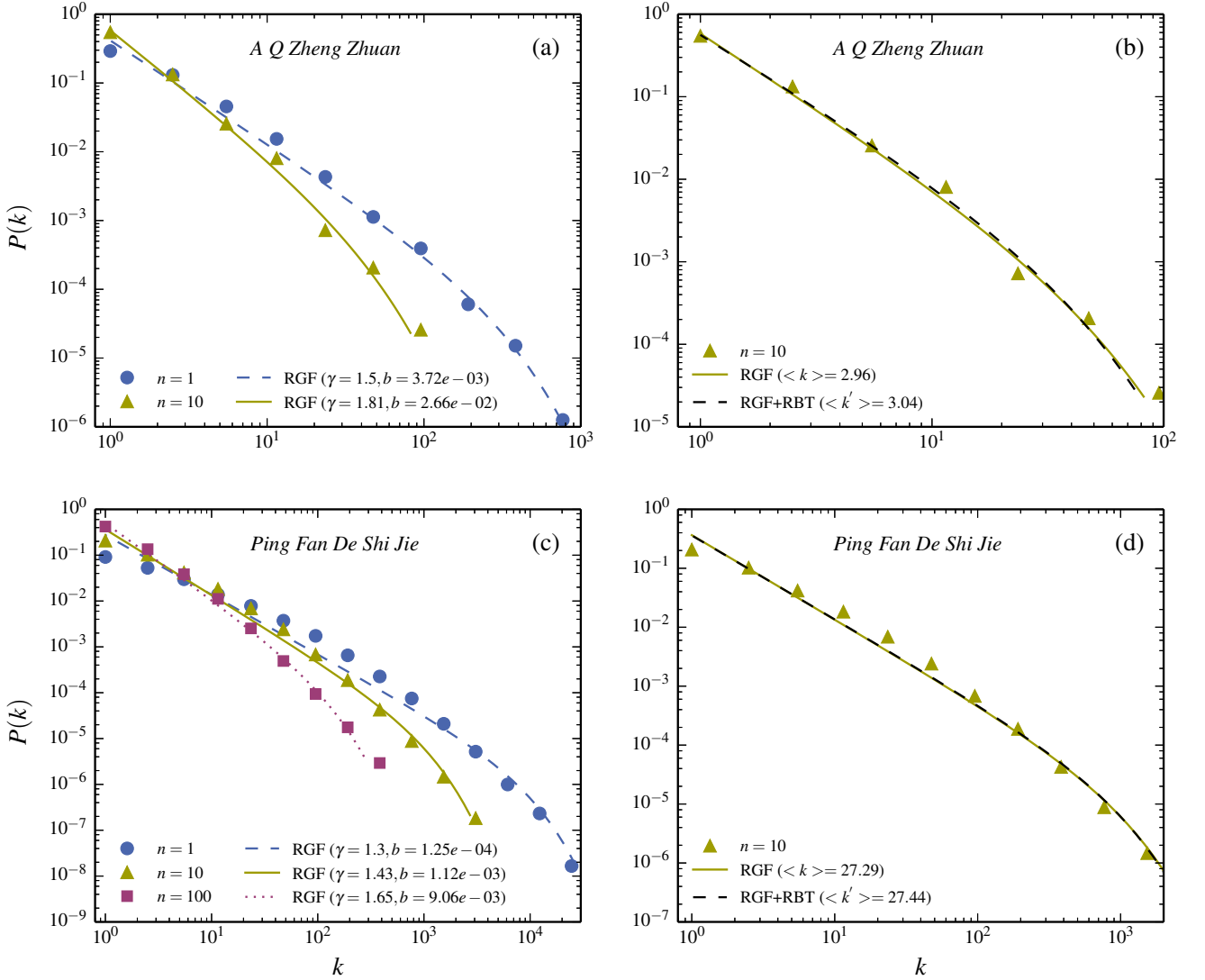}
\caption{
{\bf Size dependence of novels written in Chinese characters.} The same two novels as in figure \ref{fig:word_character} are divided into parts. The frequency distribution of a full novel is compared to the one of a part. (a) $P(k)$ for \textit{A Q Zheng Zhuan} (filled dots) is compared to the distribution for a typical 10$^{th}$-part (filled triangles). {Here the word \textit{typical} means an average distribution obtained by taking many different 10$^{th}$ with different starting points.} These two functions have quite different shapes. However, the shapes of both are equally well predicted by RGF (curves with dashed and full lines). (b) The distribution of the 10$^{th}$-part, which can to very good approximation be trivially obtained from the full book by just \textit{randomly} removing 90\% of the words from the full book. This corresponds to the dashed curve which is almost identical to the RGF-prediction and both correspond very well to the data. (c-d) The same features for the novel \textit{ Ping Fan De Shi Jie}. Note that the 10$^{th}$-part agrees better with RGF than the full novel.    
}
\label{fig:parts}
\end{figure*}

\begin{figure*}[!htbp]
\includegraphics[width=1\textwidth]{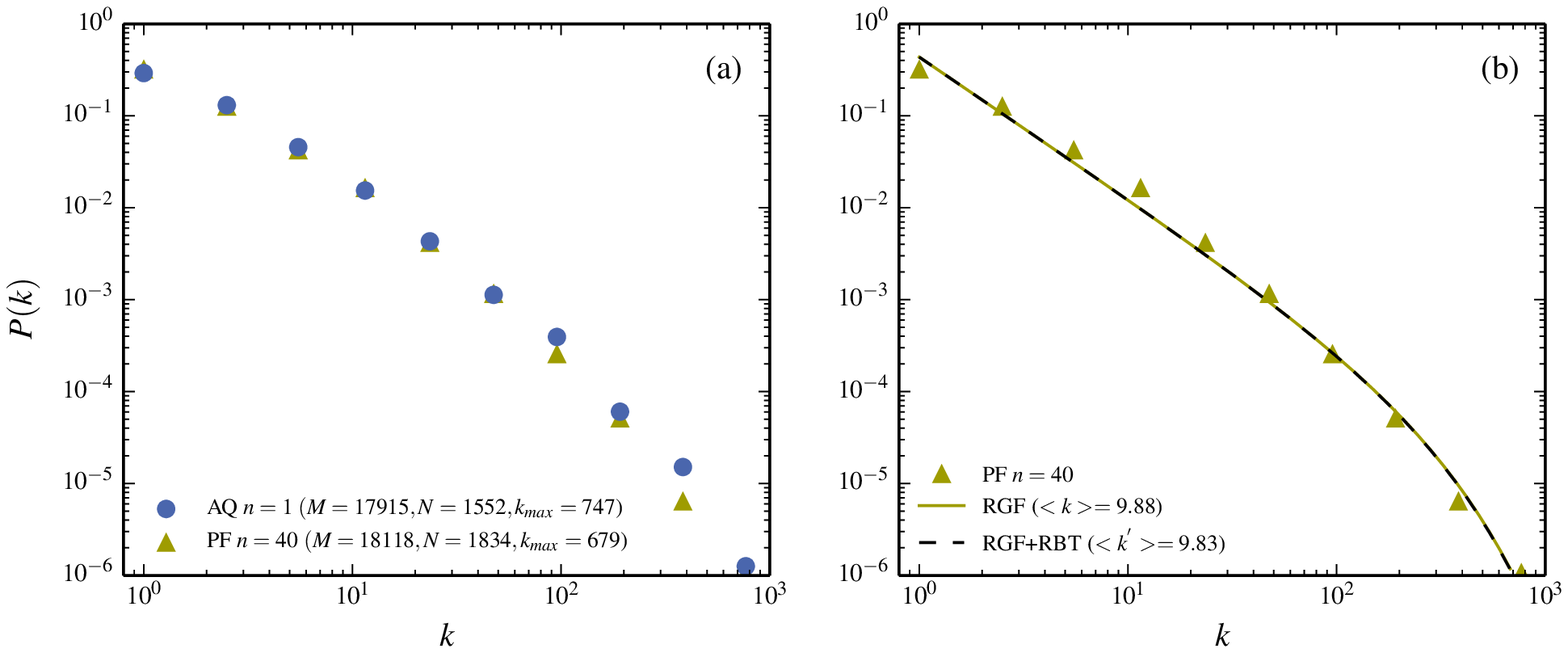}
\caption{
{\bf Comparison between two different texts of approximately equal length written by different authors in Chinese characters.} (a) \textit{A Q Zheng Zhuan} (filled dots) is compared to the 40$^{th}$ part of \textit{Ping Fan De Shi Jie} (filled triangles). Note that the two data sets almost completely overlap. This means the difference in the frequency distribution between \textit{A Q Zheng Zhuan} and \textit{Ping Fan De Shi Jie} is just caused by the difference in length of the two novels. Furthermore (b) illustrates that this length difference is rather trivial because it just the frequency distribution you get when randomly removing 97.5\% of the words from \textit{Ping Fan De Shi Jie} (dashed curve). 
}
\label{fig:length}
\end{figure*}

\section*{Results}

\subsection*{RGF and size transformation for Chinese texts}

Figure \ref{fig:word_character} shows that the data for the novel \textit{A Q Zheng Zhuan} is well described by the {neutral-model} prediction provided by RGF. This implies that the frequency distribution {of both words and characters} is to large extent directly determined by the ``state"-variable triple $(M,N,k_{max})$. At first sight this might appear surprising because the development of a spoken language and its written counterpart is a long and intricate process. However, in statistical physics this type of emergent simple properties from a complex system is well established. A well-known example is the ideal gas law $P=NT/V$ which predicts the pressure, $P$, that an ideal gas inside a closed container exerts on the walls from the three ``state"-variables $(N,V,T)$, where $N$ the number of gas particles, $V$ is the volume of the container and $T$ is the absolute temperature of the gas. Yet each gas particle follows its own deterministic trajectory including collisions with other particles and the walls. Since the number of particles is enormous it is in practice impossible to predict the outcome by deterministically following what happens in time to all the particles. The emergence of the simple ideal gas law stems from the fact that, with an enormous number of possibilities, the actual one is very likely to be close to the most likely outcome, assuming that all possibilities are equally likely. The basis for the maximum entropy principle in the present context is precisely the assumption that all distinct possibilities are equally likely. 

A crucial point is that, provided RGF does give a good description of the data, this means that it is the deviations between the data and the RGF-prediction which may carry interesting system-specific information. From this perspective Zipf's law is just an approximation of the RGF \textit{i.e.} the straight line in Fig. \ref{fig:AQ} should be regarded as an approximation of the dashed curve. It follows that the deviation between Zipf's law and the data does not reflect any characteristic property of the underlying system \cite{baek11}.

\begin{table*}[!htbp]
\caption{ 
{\bf Data and RGF-predictions.}
 Two Chinese novels are used as empirical data \textit{i.e.} \textit{A Q Zheng Zhuan} (AQ for short) written by Xun Lu and \textit{Ping Fan De Shi Jie} (PF for short) by Yao Lu. For each book we first remove punctuation marks and numbers from the texts, then count the Chinese characters one by one and finally get the characters frequency results. In Chinese language the words are not separated by spaces, so we use a word segmenter, Jieba (https://github.com/fxsjy/jieba), to extract words from Chinese texts. The RGF-prediction is given in the form $P(k)=A^{'}\exp(-bk)/k^\gamma$. This means that the RGF-theory transforms the data-triple $(M,N,k_{max})$ into the prediction triple $(\gamma,b,A^{'})$.}
\begin{tabular*}{18cm}{@{\extracolsep{\fill}}|l|r|r|r|r|r|r|}
\hline
Data Set  & $M$ & $N$ & $k_{max}$ & $\gamma$ & $b$ & $A^{'}$ \\
\hline
AQ (characters) & 17,915 & 1,552 & 747 & 1.50 & 3.72e-03 & 0.416 \\ 
AQ ($10^{th}$ part) & 1,791 & 605 & 82 & 1.81 & 2.66e-02 & 0.597 \\ 
AQ ($10^{th}$ part) RBT & 1,791 & 589 & 75 & 1.73 & 3.28e-02 & 0.581 \\ 
AQ (words) & 11,674 & 3,088 & 718 & 2.04 & 1.04e-03 & 0.624 \\ 
PF (characters) & 724,728 & 3,681 & 26,969 & 1.30 & 1.25e-04 & 0.275 \\ 
PF ($10^{th}$ part) & 72,472 & 2,655 & 2,715 & 1.43 & 1.12e-03 & 0.365 \\ 
PF ($10^{th}$ part) RBT & 72,472 & 2,641 & 2,695 & 1.42 & 1.18e-03 & 0.360 \\ 
PF ($40^{th}$ part) & 18,118 & 1,834 & 679 & 1.54 & 4.09e-03 & 0.437 \\ 
PF ($40^{th}$ part) RBT & 18,118 & 1,842 & 675 & 1.54 & 4.12e-03 & 0.437 \\ 
PF ($100^{th}$ part) & 7,247 & 1,335 & 273 & 1.65 & 9.06e-03 & 0.501 \\ 
PF (words) & 438,344 & 33,826 & 26,187 & 1.84 & 4.55e-05 & 0.548 \\
\hline
\end{tabular*}
\label{tb:1}
\end{table*}

Following this line of argument, it is essential to establish just how well the RGF does describe the data. Figure \ref{fig:word_character}(a) gives such a quality test: if all that matters is the ``state"-variables $(M,N,k_{max})$, then one could equally well translate the same novel from Chinese characters to words. As seen in Fig. \ref{fig:word_character}(a), the word-frequency distribution for the novel \textit{A Q Zheng Zhuan}  is completely different from the character-frequency and also the ``state"-variables are totally different (see Table \ref{tb:1} for ``state"-variables and RGF prediction values). Yet according to RGF the change in shape only depends on the value of the ``state"-variables and not if they relate to characters or words. As seen from Fig. \ref{fig:word_character}(a), RGF does indeed give a very good description in both cases. 

The translation of \textit{A Q Zheng Zhuan} from characters to words is in itself an example of a deterministic process. Yet, as illustrated in Fig. \ref{fig:word_character}(a), it is a complicated process in the sense that the resulting word-frequency distribution, through RGF, can be obtained to very good approximation without having any knowledge about the actual deterministic translation-process! This can again be viewed as a case when complexity results in simplicity.

Figure \ref{fig:word_character}(b) gives a second example for a longer novel, \textit{ Ping Fan De Shi Jie} by Lu Yao (about 40 times as many characters as \textit{A Q Zheng Zhuan}, see Table  \ref{tb:1}). In this case the word-frequency is very well accounted for by RGF. Note that in this particular case the Zipf's law prediction agrees very well with both the RGF-prediction and the data (Zipf's law is a straight line with slope -2 in figures \ref{fig:word_character}(a) and (b)). RGF also provides a reasonable approximation of the character-frequency, whereas Zipf's law fails completely for this case.
This is consistent with the interpretation that Zipf's law is just an approximation of RGF; an approximation which sometimes works and sometimes does not. However, as will be argued below, the discernible deviation between RGF and the data may reflect some specific linguistic feature.

As shown above, the shape of the frequency curve for a given text changes when translating between characters and words and this change is well accounted for by the RGF and the corresponding change in ``state"-variables. This is quite similar to the change of shape when more generally translating a novel to different languages. This analogy is demonstrated on the basis of the Russian short story \textit{The Man in a Case}  by A. Chekhov and its translations into English words and Chinese characters. As shown in Fig.\ref{fig:trans}(a), the respective RGF-predictions match the corresponding frequency distributions very well. The same is true for the English novel \textit{The Old Man and the Sea} by E. Hemmingway (compare Fig.\ref{fig:trans}(b)). These findings confirm that the information contained in the triple ($M$,$N$,$k_{max}$) is sufficient to describe the frequency distribution of the fundamental entities of a written language, independent if those are words or characters in Chinese and irrespective of the underlying language.

{In order to gain further insight into what causes the difference in word-frequency and character-frequency of a text written in Chinese one can compare text-parts of different lengths from a given novel. As described in \cite{bern11}, text-parts of different length of a novel have different frequency distributions.}  For example if you start from \textit{A Q Zheng Zhuan} and take an $10^{th}$-part, then the shape changes, as shown in Fig. \ref{fig:parts}(a). According to RGF this new shape should now to good approximation be directly predicted from the new ``state" $(M/10,N',k'_{max})$ (see Table  \ref{tb:1} for the precise values) As seen in Fig. \ref{fig:parts}(b) this is to good approximation the case. As explained in {\bf Methods} and can be verified from Table  \ref{tb:1}, $k'_{max}\approx k_{max}/10$. One may then ask if the transformation from $N$ to $N'$ involves some system specific feature. In order to check this one can compare the process of taking an $n^{th}$-part of a text with the process of randomly deleting characters until only a $n^{th}$-part of them remains. This latter process is a trivial statistical transformation described in {\bf Methods} under the name RBT (Random-Book-Transformation). Figure \ref{fig:parts}(b) also shows the predicted frequency distribution obtained from the ``state"-variable triple $(M',N',k_{max}')$ \textit{derived} from RBT and used as input in RGF. (The actual RBT-derived value for $N'$ is given in Table  \ref{tb:1}). The close agreements signal that the change of shape due to a reduction in text length, to large extent, is a general totally system-independent feature. Figure \ref{fig:parts}(c) shows the change of the frequency-distribution, when taking parts of the longer novel \textit{ Ping Fan De Shi Jie} written in characters and Fig. \ref{fig:parts}(d) compares the parts with the RGF-prediction, as well as with the combined RGF+RBT-prediction. The conclusion is that the change of shape carries very little system specific information.

By comparing Figs. \ref{fig:word_character}(a) and (b), one notices that whereas RGF gives a very good account of the shorter novel \textit{A Q Zheng Zhuan}, there appears to be some deviation for the longer novel \textit{ Ping Fan De Shi Jie}. In Fig. \ref{fig:length}(a) we compare a $40^{th}$ part of \textit{Ping Fan De Shi Jie} with the full length of \textit{A Q Zheng Zhuan}. As seen from Fig. \ref{fig:length}(a) the two texts have very closely the same character-frequency distribution. From the point of view of RGF, it would mean that the ``state"-variables $(M, N, k_{max})$ are closely the same. This is indeed the case, as seen in Table  \ref{tb:1} and from the direct comparison with RGF in Fig. \ref{fig:length}(b). \textit{ Ping Fan De Shi Jie} and its partitioning suggest a possible specific additional feature for written texts: a deviation from RGF for longer texts, which becomes negligible for shorter. In the following section we suggest what type of feature this might be.

\subsection*{Systematic deviations, information loss and multiple meanings of words}

As suggested in the previous section, the clearly discernible deviation in Fig. \ref{fig:word_character}(b) between the character-frequency distribution for the data and the RGF-prediction in case of  \textit{Ping Fan De Shi Jie} could be a systematic difference. The cause of this deviation should then be such that it becomes almost undetectable for a $40^{th}$-part of the same text, as seen in Fig. \ref{fig:length}(b). 

We here propose that this deviation is caused by the specific linguistic feature that a written word can have more than one meaning. {Let} us start from an English alphabetic text. A word is then defined as a collection of letters partitioned by blanks (or other partitioning signs). Such a written word could then \textit{within} the text have more than one meaning. Multiple meanings here means that a word in a dictionary is listed to have several meanings \textit{i.e.} a written word may consists of a group of words with different meanings. We will call the members of these under-groups primary words. So in order to pick a distinct primary word, you first have to pick a written word and then one of its meanings within the text.
It follows that the longer the text is, the larger the chance that several meanings of a written word appear in the text. Our explanation is based on an earlier proposed specific linguistic feature that a more frequently written word occurring in the text, has a tendency of having more meanings \cite{zipf45,reder74,cancho05,manin08}. This means that a written word which occurs $k$ times in the text on the average consists of a larger number of primary words than a written word which occurs fewer times. Thus if the text consists of $N(k)$ written words which occur $k$ times in the text, then the average number of primary words is $N_P(k)=N(k)f(k)$ where $f(k)$ describes how the number of multiple meanings depend on the frequency of the written word. {In the case of texts written with Chinese characters, it is, as explained the introduction, the characters are the elementary entities carrying individual meanings and hence play the role of words.}

\begin{figure*}[!htbp]
\includegraphics[width=1\textwidth]{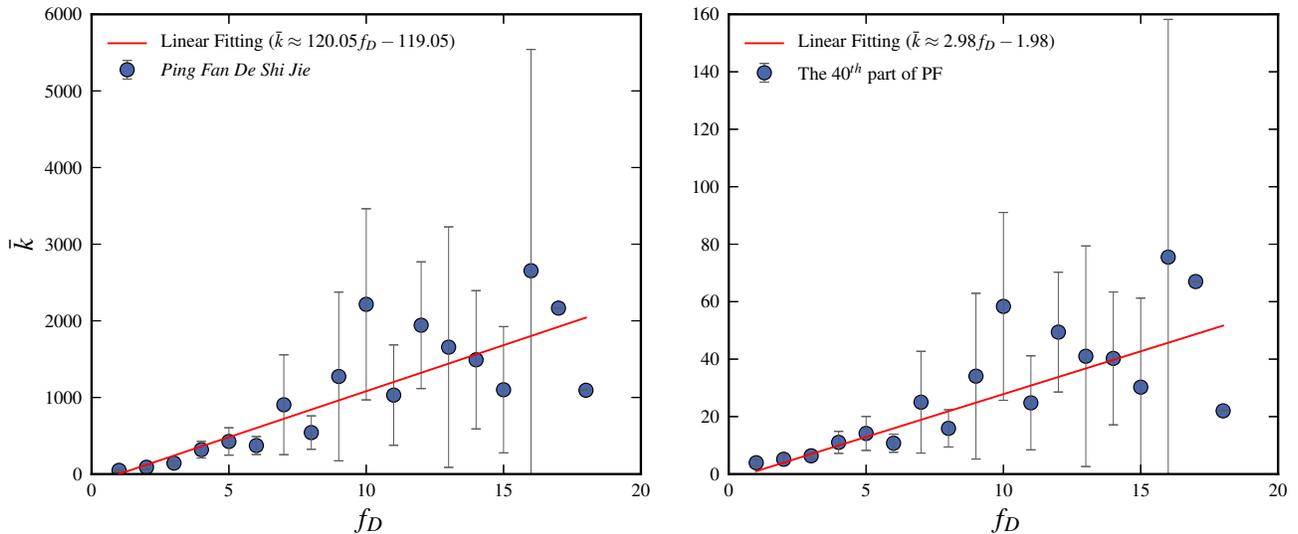}
\caption{
{\bf The average frequency $\bar{k}$ for the occurrence of a Chinese character in a given text is plotted against its number of multiple dictionary meanings $f_D$.} The Chinese character dictionary \textit{Xinhua Dictionary}, $5th$ Edition is used for the number of dictionary meanings of Chinese characters. Figure (a) shows the occurrence in the novel \textit{ Ping Fan De Shi Jie} and figure (b) the occurrence for the average 40$^{th}$-part of the same novel. In both cases the trend of the functional dependence can be represented by a straight line. The linear increase $f_D\propto c'\bar{k}$ is for the full novel $c'\approx0.0083$ and for the 40$^{th}$-part $c'\approx 0.34$. The reason that $c'$ increases with decreasing size is explained in the text.  
}
\label{fig:freq}
\end{figure*}

\begin{figure*}[!htbp]
\includegraphics[width=1\textwidth]{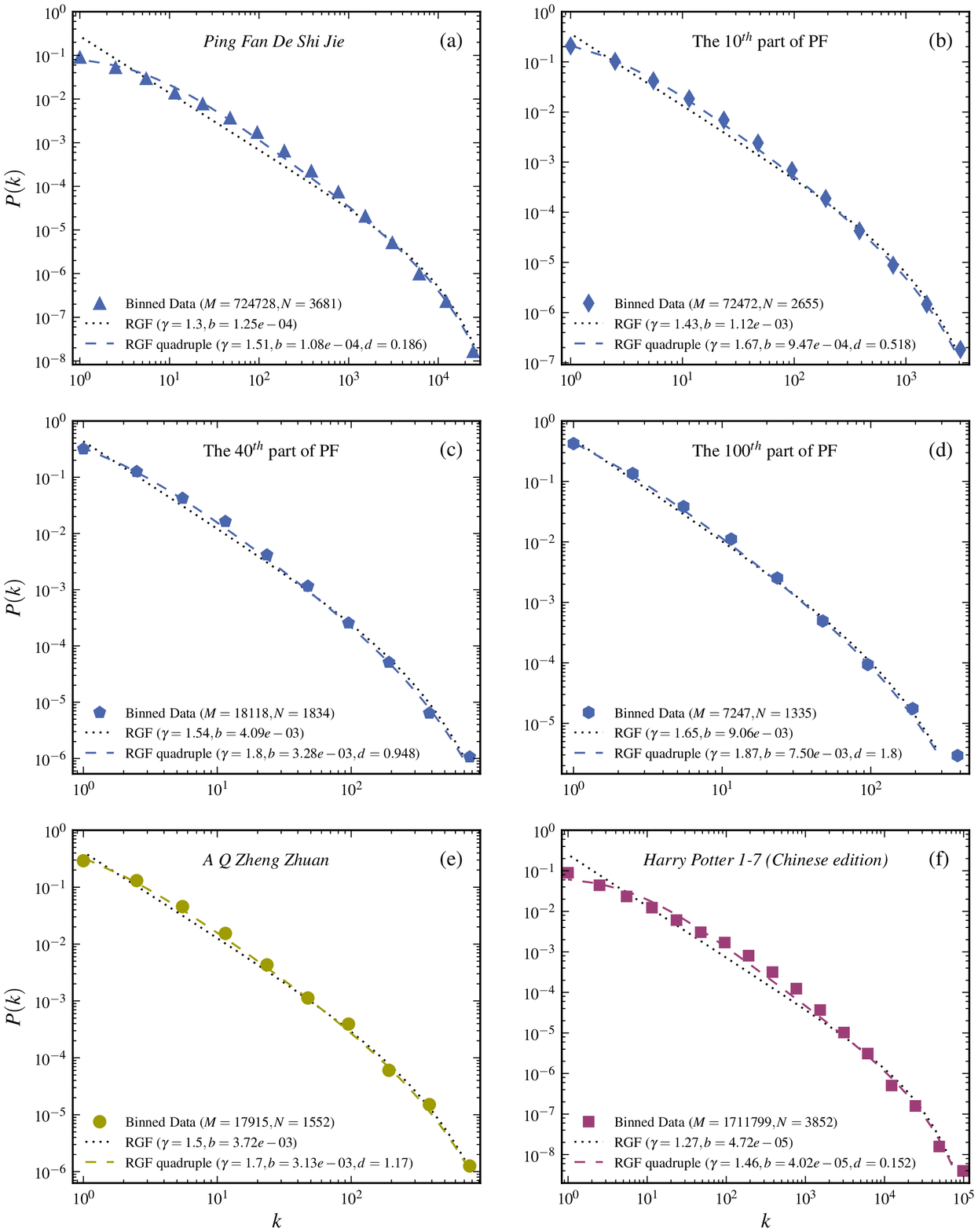}
\caption{
{\bf Test of RGF including multiple meaning constraints.} The RGF is in each case predicted from the quadruple of state variables $(M,N,k_{max},S)$. The data is from three novels in Chinese (see Table  \ref{tb:2}). The RGF predictions with multiple meaning constraint are given by the dashed curves. The RGF \textit{without} the multiple meaning constraint is predicted from the state variable triple $(M,N,k_{max})$ and corresponds to the dotted curves. Only when the multiple meaning constraint significantly improves the RGF-prediction can some specific interpretation be associated with it. As seen from the figure the significance increases with increasing length of the novel.    
}
\label{fig:rgf_mult}
\end{figure*}

It is possible to incorporate the concept of multiple meanings into a RGF-type formulation. The point to note is that the distributed entities are really the primary words/Chinese-characters and the information needed to localize a primary word/Chinese-character belonging to a written word/Chinese-character which occurs $k$ times in the text is $\log_2(kN_P(k))=\log_2(kN(k)f(k))$. We want to determine the distribution $N(k)$ taking into account that the information lost, $-\log_2(f(k))$, caused by the number of multiple multiple meanings (on the average) of a word which occurs $k$ times in the text. It follows the information which then needs to be minimized in order to obtain the maximum entropy solution is the average of $\log_2(kN(k)) -\log_2(f(k))$ or equivalently

\begin{equation}
I[N(k)]=N^{-1}\sum_k N(k) \ln(kN(k)f^{-1}(k))
\end{equation}
and following the same steps as in {\bf Methods} and \cite{baek11} this predicts the functional form 
\begin{equation}
\label{eq_primary}
P(k)=A^{'}\frac{\exp(-bk)}{(kf^{-1}(k))^{\gamma'}}.
\end{equation}

\begin{figure*}[!htbp]
\includegraphics[width=1\textwidth]{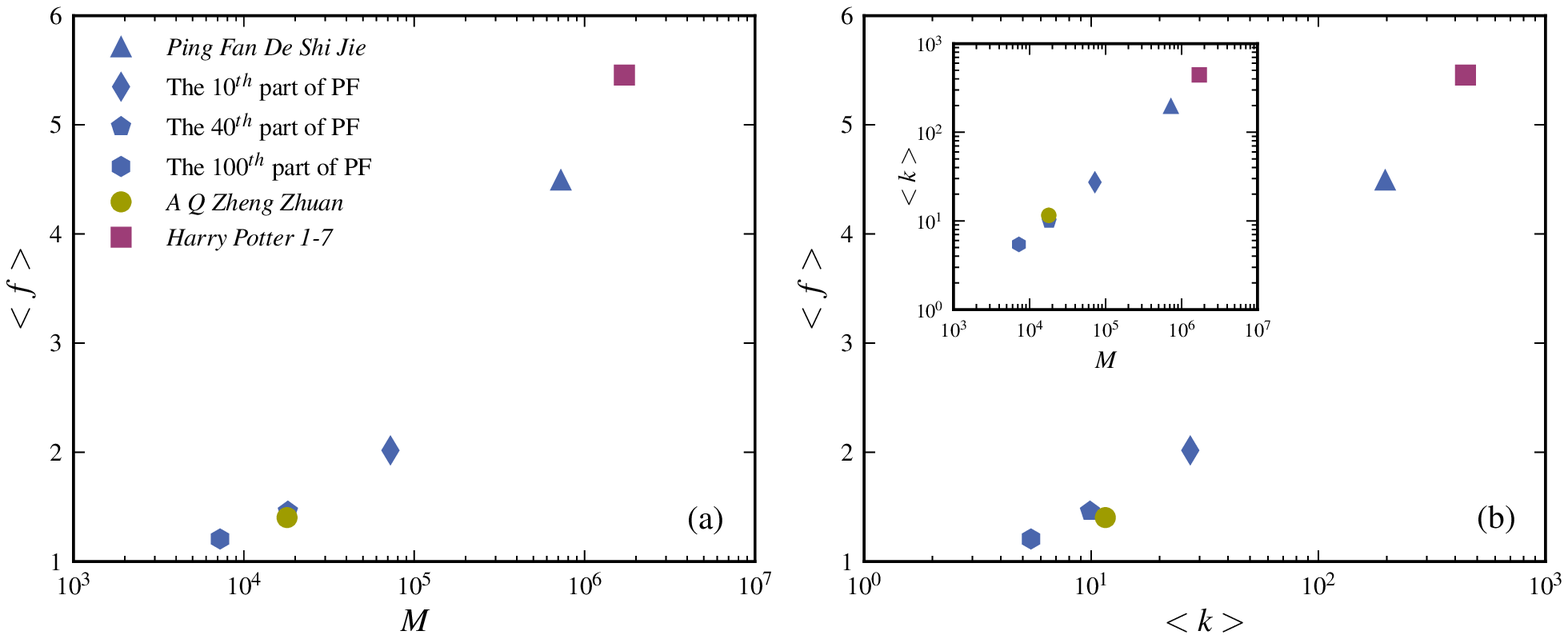}
\caption{
{\bf Consistency test of the multiple meaning model.} According to the multiple meaning model the parameter $d$ (see Table  \ref{tb:2}) should give a sensible approximative estimate of the average number of multiple meanings per character within a text $<f>$. The figure shows that $<f>$ increases with the size of the text $M$. This is consistent with the fact the number of uses of a character increases and hence the chance that more of its multiples meanings appears in the text. For the same reason $<f>$ increases with the average number of uses of a character $<k>$. In addition the chance for a larger number of dictionary meanings is larger for a more frequent character (see figure \ref{fig:freq}). The inset shows how $<k>$ increases with $M$.
}
\label{fig:f}
\end{figure*}

\begin{table*}[!htbp]
\caption{
{\bf Data and RGF-predictions including multiple meanings.}
Three Chinese novels are used as the empirical data \textit{i.e.} \textit{A Q Zheng Zhuan} written by Xun Lu, \textit{Ping Fan De Shi Jie} by Yao Lu and \textit{Harry Potter} (HP for short) volume 1 to 7 (written by J. K. Rowling and translated to Chinese by Ainong Ma \textit{et al.}). The statistics for the characters are obtained as described in table \ref{tb:1}. In this case the input quadruple $(M,N,k_{max},S)$ is transformed by the RGF-theory into the output prediction $(\gamma,b,d,A^{'})$ corresponding to the RGF-form $A^{'}\frac{\exp(-bk)}{k^{\gamma}(1+\frac{1}{dk})^{\gamma}}$.
}

\begin{tabular*}{18cm}{@{\extracolsep{\fill}}|l|r|r|r|r|r|r|r|r|r|}
\hline
Data Set & $M$ & $N$ & $k_{max}$ & $S$ & $\gamma$ & $b$ & $d$ & $A^{'}$ & $<f>$ \\
\hline
PF (characters)  & 724,728 & 3,681 & 26,969 & 5.03 &1.51 & 1.08e-04 & 0.186 & 1.314 & 4.49\\ 
PF ($10^{th}$) & 72,472 & 2,655 & 2,715 & 3.51 & 1.67 & 9.47e-04 & 0.518 & 1.275 & 2.02\\ 
PF ($40^{th}$) & 18,118 & 1,834 & 679 & 3.51 & 1.80 & 3.28e-03 & 0.948 & 1.219 & 1.46\\ 
PF ($100^{th}$) & 7,247 & 1,335 & 273 & 2.19 & 1.87 & 7.50e-03 & 1.800 & 1.013 & 1.21\\ 
AQ (characters)  & 17,915 & 1,552 & 747 & 2.81 & 1.70 & 3.13e-03 & 1.170 & 0.938 & 1.40\\ 
HP (characters)  & 1,711,799 & 3,852 & 71,262 & 5.49 & 1.46 & 4.02e-05 & 0.152 & 1.177 & 5.45\\
\hline
\end{tabular*}
\label{tb:2}
\end{table*}

Basically the specific linguistic character is that $f(k)$ is an increasing function and that $f(k=1)=1$, because a word which only occurs a single time in the text can only have one meaning within the text. The simplest approximation is then just a linear increase. Figure \ref{fig:freq} gives some support for this supposition: the average frequency, $\bar{k}(f_D)$, of Chinese characters in \textit{ Ping Fan De Shi Jie}, which have $f_D$ dictionary meanings,is plotted against $f_D$. The plot shows that the $\bar{k}(f_D)$ to fair approximation has a linear increase of the form $\bar{k}=f_D/c'-1/c'+1$ or equivalently $f_D=c'\bar{k}+1-c'$. Figure \ref{fig:freq}(a) corresponds to the full text and Fig. \ref{fig:freq}(b) to a $40^{th}$ part. Note that the slope $c'$ changes with text size. This is easily understood: shortening the text is, as explained in the previous section, basically the same as randomly removing characters. This means that a character with a smaller $k$ has a larger chance to be completely removed from the text than one with higher. But since the characters with higher frequency on average have a larger number of multiple meanings, this means that the resulting characters with low $k$ will on average have more multiple meanings. 
Also note that the \textit{dictionary} meanings and the meanings \textit{within} a text is not the same; the former is larger than the latter, but the longer the text the more equal they become. However, it is reasonable to assume that also the number of meanings \textit{within} a text follows a similar linear relationship. Next we make the further simplification by replacing the average $\bar{k}$ with just $k$ \textit{i.e.} we are ignoring the spread in frequency of characters having a specific number of meanings within the text. However, this approximation still catches the increase in meanings with frequency. We will take this linear increase as our ansatz and include a cut-off $k_c$ for large $k$, since the most frequent Chinese characters has few multiple meanings. This is a general linguistic feature, the most frequent English words, "the", has only one meaning. Thus we use the approximate ansatz $f(k) \propto k/(1+k/k_c)$. This approximation reduces the RGF functional form to
\begin{equation}
\label{eq:lin}
P(k)=A^{'}\frac{\exp(-bk)}{k^\gamma(1+\frac{1}{dk})^{\gamma}},
\end{equation}
where $d=1/k_c$. In addition to the ``state"-variable triple $(N,M,k_{max})$ we should specify an a priori knowledge of $f(k)$. The knowledge of this linguistic constraint is limited and enters through its \textit{ approximate} form $f(k)\propto k/(1+kd)$. This enables us to determine the value $d=1/k_c$ from the RFG-method by including  the value of the entropy $S$ as an additional constraint. Thus we use RGF in the form of (\ref{eq:lin}) together with the ``state"-variable quadruple $(N,M,k_{max},S)$. {This follows since the four constants $(A^{'},b,\gamma,d)$ in Eq.(\ref{eq:lin}), through RGF-formulation completely determine the quadruple $(N,M,k_{max},S)$ and vice versa.} In Fig. \ref{fig:rgf_mult} this form of extended RGF is tested on data from three novels written in Chinese characters. The corresponding ``state"-quadruples $(N,M,k_{max},S)$ are given in Table  \ref{tb:2} together with the corresponding predicted output-quadruple $(\gamma,b,k_{max},d)$. The agreement with the data is in all cases excellent (dashed curves in the Fig. \ref{fig:rgf_mult}). The dotted curves are the usual RGF-prediction based on the ``state"-triples $(M,N,k_{max})$. Note that for a 100$^{th}$-part of \textit{ Ping Fan De Shi Jie}, the usual RGF and the extended RGF agrees equally well with the data. This means that any effect of multiple meanings is in this case already taken care of by the usual RGF. However as the text size is increased to 40$^{th}$-, 10$^{th}$ part and full novel, the extended RGF agrees equally well, whereas the usual RGF-start to deviate. It is this systematic difference, which suggest that there is specific effect beyond the {neutral-model} prediction given by the usual RGF.

\begin{figure*}[!htbp]
\includegraphics[width=1\textwidth]{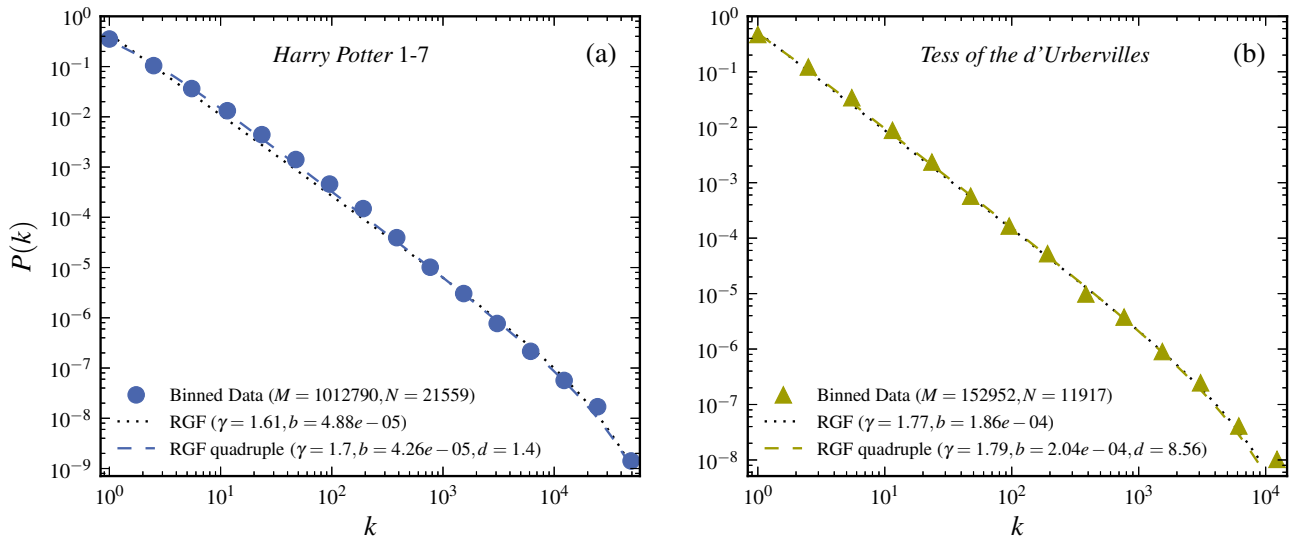}
\caption{
{\bf Test of RGF including multiple meaning constraints for English books.} The RGF is in each case predicted from
the quadruple of state variables ($M, N, k_{max} , S$). The data is from two English novels: \textit{Tess of the d'Urbervilles} written by T. Hardy and \textit{Harry Potter} volume 1 to 7 by J. K. Rowling. The RGF predictions with multiple meaning constraint are given by the dashed curves. The RGF without the multiple meaning constraint is predicted from the state variable triple ($M, N, k_{max}$ ) and corresponds to the dotted curves.
}
\label{fig:en}
\end{figure*}

Is the multiple meaning explanation sensible? To investigate this we estimate the average number of multiple meanings $<f(k)>$ using the ansatz form for $f$ including the condition that a single character can only have a single meaning in the text $f(k=1)=1$ \textit{i.e.} $f(k)=(1+d)k/(1+kd)$ together with the  obtained values of $d$ (see Table  \ref{tb:2})
   
\begin{equation}
 \sum_{k=1}N(k)f(k)N= \sum_{k=1}^{k_c}P(k)\frac{(1+d)k}{1+dk}.
\end{equation}
These estimated values for $<f>$ are given in Table  \ref{tb:2}.
Figure \ref{fig:f}(a) shows that $<f>$ increases with the text length. This is consistent with the fact that the number of uses of a character increases and hence the chance that more of its multiples meanings appears in the text. For the same reason $<f>$ increases with the average number of uses of a character $<k>$ as shown in Fig. \ref{fig:f}(b). In addition the chance for a larger number of dictionary meanings is larger for a more frequent character (see Fig. \ref{fig:freq}). Thus it appears that the connection between $<f>$ and multible meanings makes sense.

Multiple meaning is of course not a unique feature of Chinese, it is a common feature of many languages. Therefore, it is unsurprising that we can also observe systematic deviations from the RGF-prediction in other languages, such as English \cite{bern11} and Russian \cite{manin08}. However, the average meaning of English words are much less than that of Chinese character: in modern Chinese there are only about $3,500$ commonly used characters \cite{yan13} and even for a novel including more than one million of characters, the number of distinct characters involved is less than $4,000$ (see Table \ref{tb:2}); but for the same novel written in English, the number of distinct words is more than $20,000$ (see Fig. \ref{fig:en}(a)). Therefore, the systematic deviation caused by multiple meaning can be neglected for short English text, as shown in Fig. \ref{fig:en}(b). Even for a rather long text, the deviation is still very slight and, as shown in Fig. \ref{fig:en}(a), the usual RGF gives a good prediction (RGF with multiple meaning constraint incorporates more \textit{a priori} information and may consequently be expected to give a better prediction but the difference is very small). Taken together, Chinese uses a small amount of characters to describe the primary word, resulting in a high degree of multiple meanings, further leading to that the head of the character-frequency distribution (or tail of the frequency-rank distribution) deviates somewhat from the RGF-prediction. But such deviations are not special to Chinese, as we have demonstrated in Fig. \ref{fig:en}, it is just more pronounced in Chinese than in some other languages.

\section*{Discussion}

\begin{figure*}[!htbp]
\includegraphics[width=1\textwidth]{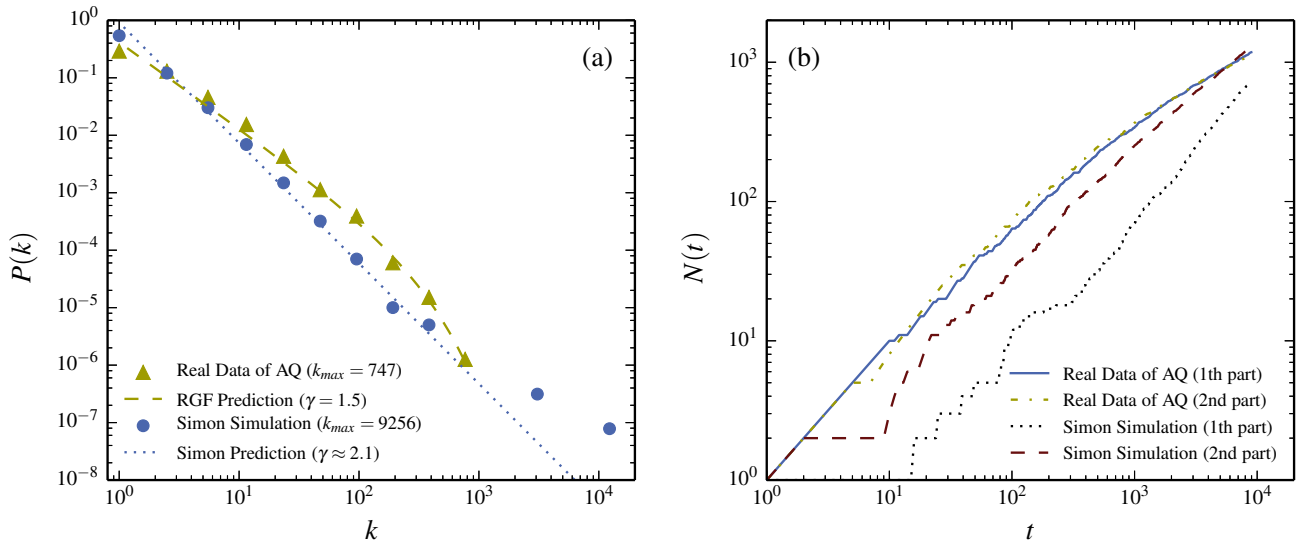}
\caption{
{\bf Test of the Simon model.} (a) The data (solid triangles) together with the RGF-prediction (dashes curve) for \textit{A Q Zheng Zhuan} in Chinese characters. The Simon model with the same $M$ and $N$ are given by the solid dots and the Simon prediction for infinite $M$ by the dotted line. Note that the most common character appears $9,256$ times for the Simon model which is about 50\% of the total number of characters. This is completely unrealistic for a sensible language (the most common character in Chinese is about 4\% and the most common word in English ``the" is also about 4\%). Figure (b) shows that the frequency distribution for Simon model is not translation invariant: For a real novel the word frequency distribution of the first half of the novel is to good approximation the same as the second. The data for the novel \textit{A Q Zheng Zhuan} in Chinese characters illustrates this (full drawn and short dashed curves in the figure). However for the Simon model the frequency distribution depends on which part you take (long dashed- and dotted curves in the figure).
}
\label{fig:simon}
\end{figure*}

The view taken in the present paper is somewhat different and heretical compared to a large body of earlier work \cite{zipf32, zipf35,zipf49,mand53,li92,baayen01,cancho03,mont01,simon55,kanter95,doro01,zanette05,wang05,masucii06,cattuto06,lu13}. First of all we argue that Zipf's law is not a good starting point, when trying to extract information from word/character frequency distributions. Our starting point is instead a {neutral-model} containing a minimal
\textit{a priori} information about the system. From this minimal information, the frequency distribution is predicted through a maximum entropy principle. The minimal information consists of the ``state"-variable triple $(M,N,k_{max})$ corresponding to the \textit{(total number of- , number of different- , maximum occurrence of most frequent-)} word/character, respectively. The shape of the distribution is entirely determined by the triple $(M,N,k_{max})$. Within this RGF-approach, Zipf's-law (or any other power law with an exponent different from the Zipf's law exponent) distribution only results for seemingly accidental triples of  $(M,N,k_{max})$. The first question is then if these Zipf's law triples are really accidental or if they carry some additional information about the system. According to our findings
there is nothing special about these power-law cases. First of all in the examples discussed here, Zipf's law is in most cases not a good approximation of the data, whereas the RGF-prediction in general gives a very good account of all the data \textit{including} the rare cases when the distribution is close to a Zipf's law. Second, translating a novel between languages, or between words and Chinese characters, or taking parts of the novel, all  changes the triple $(M,N,k_{max})$. This means that the shape of the distribution changes, such that if it happened to be close to a Zipf's law before the change, it deviates after. Furthermore, in the case of taking parts of a novel, the change in the triple $(M,N,k_{max})$ is to large extent trivial, which means that there is no subtle constraint for preferring special values of $(M,N,k_{max})$. All what this leads up to is that the distributions you find in word/character frequencies are very general and apply to any system which can be similarly described in terms of the triple $(M,N,k_{max})$ as discussed in  \cite{baek11, bokma13}. From this point of view the word/character frequency carries little specific information about languages.

In a wider context, this generality and lack of system-dependence was also expressed in  \cite{bokma13} as: \textit{...we can safely exclude the possibility that the processes that led to the distribution of avian species over families also wrote the United States' declaration of independence, yet both are described by RGF,}
and earlier and more drastically by Herbert Simon in  \cite{simon55}: \textit{No one supposes that there is any connection between horse-kicks suffered by soldiers in the German army and blood cells on a microscopic slide other than that the same urn scheme provides a satisfactory abstract model for both phenomena.} The urn scheme used in the present paper is the maximum entropy principle in the form of RGF.

Herbert Simon's own urn model is called the Simon model \cite{simon55}. The problem with the Simon model in the context of written text is that it does presume a specific relation between the parameters of the ``state"-triple $(M,N,k_{max})$: for a text with a given $M$ and $N$, the Simon model \textit{predicts} a $k_{max}$. This value of $k_{max}$ is quite different from the ones describing the real data analyzed here. For example in case of the ``state" triple for \textit{A Q Zheng Zhuan} in Chinese characters the values of $M$and $N$ are $17,915$ and $1,552$, respectively (see Table  \ref{tb:1}) and the Simon  model predicts $k_{max}=9,256$ and $P(k)$ in the form of a power law given by $\propto 1/k^{2.1}$. Thus the most common character accounts for about 50\% of the total text, which does not correspond to any realistic language. Figure \ref{fig:simon}(a) compares this Simon model result with the real data, as well as with the corresponding RGF-predictions. You could perhaps imagine that you in each case could modify the Simon model so as to produce the correct ``state"-triple. However, even so a modified Simon models will anyway have a serious problem, as discussed in  \cite{bern11}: if you take a novel written by the Simon stochastic model and divide it into two equally sized parts, then the first part has a quite different triple $(M/2,N_{1/2},k_{max}/2)$ than the second. Yet both parts of a real book are described by the same ``state"-variable triple. This means that the change in shape of the distribution by partitioning cannot be correctly described within any stochastic Simon-type model. 

From the point of view of the present approach, the fact that the data is very well described by the RGF-model gives a tentative handle to get one step further: since RGF is a {neutral-model} prediction, the implication is that any systematic deviations between the data and the RGF-prediction might carry additional specific information about the system. Such a deviation was shown to become more discernable the longer the text written in Chinese characters is. The multiple meaning of Chinese characters was suggested as an explanatory factor of this phenomenon. This is based on the notion that characters/words used with larger frequency have a tendency to have more multiple meanings within a text. Some supports for this was gained be comparing to the dictionary meanings of a Chinese character. It was also argued that this tendency of more multiple meanings could be entered as an additional constraint within the RGF-formulation. Comparison with data suggested that this is indeed a sensible contender for an explanation.

Our view is that the neutral-model provided by RGF provides a useful starting point for extracting information from word/character distributions in texts. It has the advantage, compared to most other approaches, in that it actually predicts the real data from a very limited amount of \textit{a priori} information. It also has the advantage of being a general approach which can be applied to a great variety of different systems.

\section*{Acknowledgments}
Economic support from IceLab is gratefully acknowledged.


\begin{thebibliography}{100}
 \bibitem{singh2000}
Singh S (2000) The Code Book. New York,USA: Random House.
\bibitem{estroup16}
Estroup JB (1916)  Les Gammes St\'enographiques (4th edn). Paris: Institut Stenographique de France.
\bibitem{zipf32}
Zipf GK (1932)  Selective Studies of the Principle of Relative Frequency in Language. Cambridge, MA: Harvard University Press.
\bibitem{zipf35}
Zipf GK (1935)  The Psycho-Biology of Language: an Introduction to Dynamic Philology. Boston, MA: Mifflin.
\bibitem{zipf49}
Zipf GK (1949)  Human Bevavior and the Principle of Least Effort. Reading, MA: Addison-Wesley.
\bibitem{mand53}
Mandelbrot B (1953)  An Informational Theory of the Statistical Structure of Languages. Woburn, MA: Butterworth.
\bibitem{li92}
Li W (1992) Random texts exhibit Zipf's-law-like word frequency distribution.  IEEE T Inform Theory   38: 1842
\bibitem{baayen01}
Baayen RH (2001)   Word Frequency Distributions.  Dordrecht, Netherlands: Kluwer Academic.
\bibitem{cancho03}
i Cancho RF, Sol\'e  RV (2003) Least effort and the origins of scaling in human language.  Proc Natl Acad Sci USA   100: 788
\bibitem{mont01}
Montemurro MA (2001) Beyond the Zipf-Mandelbrot law in quantitative linguistics.  Physica A   300: 567
\bibitem{simon55}
Simon H (1955) On a class of skew distribution functions.  Biometrika   42: 425
\bibitem{kanter95}
Kanter I, Kessler DA (1995) Markov processes: linguistics and Zipf's Law.   Phys Rev Lett    74: 4559
\bibitem{doro01}
Dorogovtsev SN, Mendes JFF (2001) Languague as an evolving word web.  Proc R Soc Lond B   268: 2603
\bibitem{zanette05}
Zanette DH, Montemurro MA (2005) Dynamics of text generation with realistic Zipf's distribution.  J Quant Linguistics 12: 29
\bibitem{wang05}
Wang DH, Li MH, Di ZR (2005) Ture reason for Zipf's law in language.  Physica A   358: 545
\bibitem{masucii06}
Masucci A, Rodgers G (2006)  Networks properties of written human language.  Phys Rev E    74: 26102
\bibitem{cattuto06}
Cattuto C, Loreto V, Servedio VDP (2006) A Yule-Simon process with memory.  Europhys Lett   76: 208
\bibitem{lu13}
L\"u L, Zhang Z-K, Zhou T (2013) Deviation of Zipf's and Heaps' laws in human languages with limited dictionary sizes.  Sci Rep   3: 1082

\bibitem{zhao90} Zhao KH (1990) Physics nomenclature in China.  Am J Phys  58: 449

\bibitem{rous92} Rousseau R, Zhang Q (1992) Zipf's data on the frequency of Chinese words revisited.  Scientometrics.  24: 201

\bibitem{shtrik94} Shtrikman S (1994) Some comments on Zipf's law for the Chinese language.  J Inf Sci  20: 142

\bibitem{ha00} Ha LQ, Sicilia-Garcia EI, Ming J, Smith FJ (2000) Extension of Zipf's law to words and phrases.  Proceedings of the 19th international conference on Computational linguistics  1: 1

\bibitem{bern10}
Bernhardsson S, da Rocha LEC, Minnhagen P (2010) Size dependent word frequencies and the translational invariance of books.  Physica A   389: 330
\bibitem{bern11}
Bernhardsson S, da Rocha LEC, Minnhagen P (2009) The meta book and size-dependent properties of written language.  New J Phys   11: 123015
\bibitem{miller57}
Miller GA (1957) Some effects of intermittance silence.  Am J Psychol 70: 311
\bibitem{bern11b}
Bernhardsson S, Baek SK, Minnhagen P (2011) A paradoxical property of the monkey book.  J Stat Mech   7: PO7013
\bibitem{baek11}
Baek SK, Bernhardsson S, Minnhagen P (2011) Zipf's law unzipped.  New J Phys   13: 043004
\bibitem{bokma13}
Bokma F, Baek SK, Minnhagen P (2013) 50 years of inordinate fondness.  Syst biol syt067


\bibitem{clauset09}
Clauset A, Shalizi CR and Newman MEJ (2009) Power-law distributions in empirical data. SIAM Rev  51: 661
\bibitem{visser13}
Visser M (2013) Zipf’s law, power laws and maximum entropy.  New J Phys   15: 043021

\bibitem{footnote} $P(k)$ is the distribution from which the data is drawn and the entropy $S$ is a functional of $P(k)$. Thus in order to obtain the functional form for fixed $S$ you do not need any \textit{a priori} knowledge of $S$ from the data. Once you have the functional form you are free to make any choice of how to relate it to the data.

\bibitem{lee12}
Lee SH, Bernhardsson S, Holme P, Kim BJ, Minnhagen P (2012) Neutral theory of chemical reaction networks.  New J Phys 14: 033032
\bibitem{baek11b}
Baek SK, Minnhagen P, Kim BJ (2011) The ten thousand Kims.  New J Phys   13: 073036
\bibitem{zipf45}
Zipf GK (1945) The meaning-frequency relationship of words.  J Gen Psychol 33: 251
\bibitem{reder74}
Reder L, Anderson JR, Bjork RA (1974) A semantic interpretation of encoding specificity.  J Exp Psychol 102: 648
\bibitem{cancho05}
i Cancho RF (2005) The variation of Zipf's law in human language.  Eur Phys J B   44: 249
\bibitem{manin08}
Manin DY (2008) Zipf's law and avoidance of excessive synonymy.  Cognitive Sci 32: 1075
\bibitem{yan13}
Yan X, Fan Y, Di Z, Havlin S, Wu J (2013) Efficient learning strategy of Chinese characters based on network approach.  PLoS ONE   8: e69745
 \end{thebibliography}
\end{document}